\documentclass[manuscript]{acmart}

\usepackage[normalem]{ulem}

\AtBeginDocument{%
  \providecommand\BibTeX{{%
    \normalfont B\kern-0.5em{\scshape i\kern-0.25em b}\kern-0.8em\TeX}}}

\setcopyright{none}
\copyrightyear{2023}
\acmYear{2023}
\acmDOI{XXXXXXX.XXXXXXX}

\acmConference[HAIGEN '23 Workshop at IUI '23]{Make sure to enter the correct
  conference title from your rights confirmation emai}{March 27-31, 2023}{Sydney, NSW, Australia}
\begin{document}

\title{Toward General Design Principles for Generative AI Applications}


\author{Justin D. Weisz}
\orcid{0000-0003-2228-2398}
\affiliation{
    \institution{IBM Research AI}
    \city{Yorktown Heights}
    \state{NY}
    \country{USA}
}
\email{jweisz@us.ibm.com}

\author{Michael Muller}
\orcid{0000-0001-7860-163X}
\affiliation{
    \institution{IBM Research AI}
    \city{Cambridge}
    \state{MA}
    \country{USA}
}
\email{michael_muller@us.ibm.com}

\author{Jessica He}
\orcid{xxxx-xxxx-xxxx-xxxx}
\affiliation{
    \institution{IBM Research AI}
    \city{Seattle}
    \state{WA}
    \country{USA}
}
\email{jessicahe@ibm.com}

\author{Stephanie Houde}
\orcid{0000-0002-0246-2183}
\affiliation{
    \institution{IBM Research AI}
    \city{Cambridge}
    \state{MA}
    \country{USA}
}
\email{Stephanie.Houde@ibm.com}

\renewcommand{\shortauthors}{Weisz et al. 2023}

\newcommand{\MM}[1]{\textcolor{magenta}{MM: #1}}
\newcommand{\SH}[1]{\textcolor{cyan}{SH: #1}}
\newcommand{\JW}[1]{\textcolor{red}{JW: #1}}
\newcommand{\JH}[1]{\textcolor{blue}{JH: #1}}

\begin{abstract}
  Generative AI technologies are growing in power, utility, and use. As generative technologies are being incorporated into mainstream applications, there is a need for guidance on how to design those applications to foster productive and safe use. Based on recent research on human-AI co-creation within the HCI and AI communities, we present a set of seven principles for the design of generative AI applications. These principles are grounded in an environment of \emph{generative variability}. Six principles are focused on \emph{designing for} characteristics of generative AI: multiple outcomes \& imperfection; exploration \& control; and mental models \& explanations. In addition, we urge designers to design \emph{against} potential harms that may be caused by a generative model's hazardous output, misuse, or potential for human displacement. We anticipate these principles to usefully inform design decisions made in the creation of novel human-AI applications, and we invite the community to apply, revise, and extend these principles to their own work.
\end{abstract}

\ccsdesc[500]{Human-centered computing~HCI design and evaluation methods}
\ccsdesc[300]{Human-centered computing~Interaction paradigms}
\ccsdesc[300]{Human-centered computing~HCI theory, concepts and models}

\keywords{generative AI, design principles, human-centered AI, foundation models}

\begin{teaserfigure}
  \centering
  \includegraphics[width=0.5\textwidth]{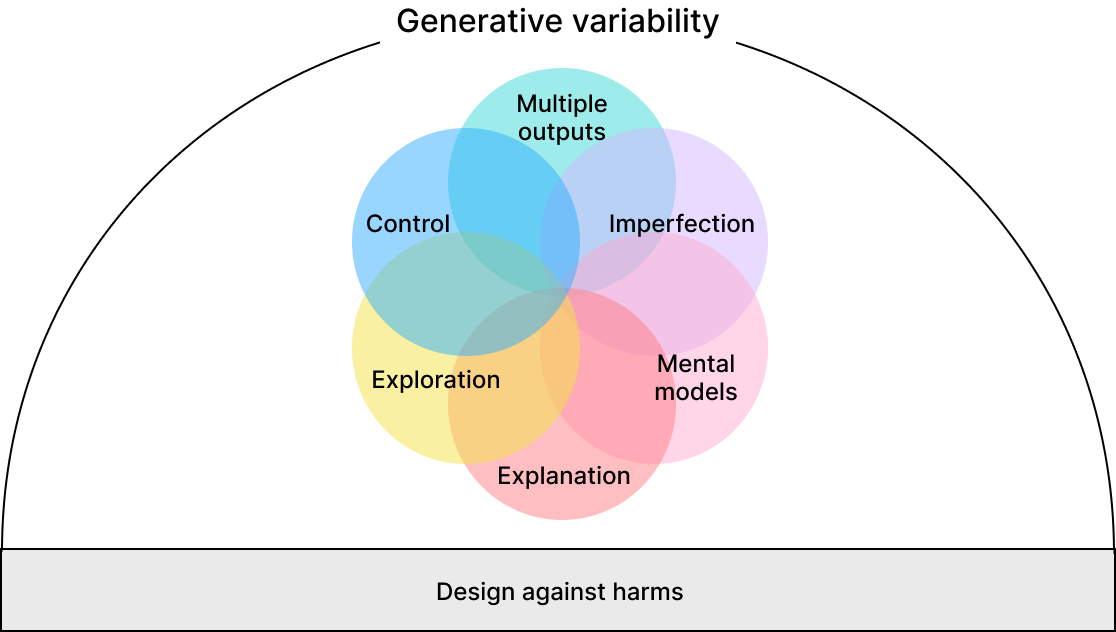}
  \caption{Seven principles for the design of generative AI systems. Six of these principles are presented in overlapping circles, indicating their relationships to each other. One principle stands alone, the directive to design against potential harms that may be caused by a generative model's output, misuse, or other harmful effects. These principles are bounded in an environment of \emph{generative variability}, in which the outputs of a generative AI application may vary in quantity, quality, character, or other characteristics.}
  \label{fig:teaser}
\end{teaserfigure}


\maketitle

\section{Introduction}

As generative AI technologies continue to grow in power and utility, their use is becoming more mainstream. Generative models, including LLM-based foundation models~\cite{bommasani2021opportunities}, are being used for applications such as general Q\&A (e.g. ChatGPT\footnote{\url{http://chat.openai.com}}), software engineering assistance (e.g. Copilot\footnote{\url{http://copilot.github.com}}), task automation (e.g. Adept\footnote{\url{http://adept.ai}}), copywriting (e.g. Jasper.ai\footnote{\url{http://jasper.ai}}), and the creation of high-fidelity artwork (e.g. DALL-E 2~\cite{ramesh2022hierarchical}, Stable Diffusion~\cite{rombach2022high}, Midjourney\footnote{\url{http://midjourney.com}}). Given the explosion in popularity of these new kinds of generative applications, there is a need for guidance on how to design those applications to foster productive and safe use, in line with human-centered AI values~\cite{shneiderman2022human}.

Fostering productive use is a challenge, as revealed in a recent literature survey by \citet{campero2022test}. They found that many human-AI collaborative systems failed to achieve positive synergy -- the notion that a human-AI team is able to accomplish superior outcomes above either party working alone. In fact, some studies have found the opposite effect, that human-AI teams produced inferior results to either a human or AI working alone~\cite{clark2018creative, buccinca2021trust, jacobs2021machine, kleinberg2021humans}.

Fostering safe use is a challenge because of the potential risks and harms that stem from generative AI, either because of how the model was trained (e.g. \cite{weidinger2021ethical}) or because of how it is applied (e.g. \cite{houde2020business, muller2022drinking}).

In order to address these issues, we propose a set of design principles to aid the designers of generative AI systems. These principles are grounded in an environment of \textbf{generative variability}, indicating the two properties of generative AI systems inherently different from traditional discriminative\footnote{Our use of the term \emph{discriminative} is to indicate that the task conducted by the AI algorithm is one of determining to which class or group a data instance belongs; classification and clustering algorithms are examples of discriminative AI. Although our use of the term \emph{discriminative} may evoke imagery of human discrimination (e.g. via racial, religious, gender identity, genetic, or other lines), our use follows the scientific convention established in the machine learning community (see, e.g., \url{https://en.wikipedia.org/wiki/Discriminative_model})} AI systems: \textbf{generative}, because the aim of generative AI applications is to \emph{produce artifacts} as outputs, rather than \emph{determine decision boundaries} as discriminative AI systems do, and \textbf{variability}, indicating the fact that, for a given input, a generative system may produce a variety of possible outputs, many of which may be valid; in the discriminative 
case, it is expected that the output of a model does not vary for a given input.

We note that our principles are meant to generally apply to generative AI applications. Other sets of design principles exist for specific kinds of generative AI applications, including \citet{liu2022design}'s guidelines for engineering prompts for text-to-image models, and advice about one-shot prompts for generation of texts of different kinds \cite{greyling:engineering, reynolds:prompt, Denny:Conversing}.
There are also more general AI-related design guidelines \cite{acm2023words, amershi2019guidelines, apple, ibm2023reid, liao2020questioning}. 

Six of our principles are presented as ``design for...'' statements, indicating the characteristics that designers should keep in mind when making important design decisions. One is presented as a ``design against...'' statement, directing designers to design \emph{against} potential harms that may arise from hazardous model outputs, misuse, potential for human displacement, or other harms we have not yet considered. The principles interact with each other in complex ways, schematically represented via overlapping circles in Figure~\ref{fig:teaser}. For example, the characteristic denoted in one principle (e.g. multiple outputs) can sometimes be leveraged as a strategy for addressing another principle (e.g. exploration). Principles are also connected by a user's aims, such as producing a singular artifact, seeking inspiration or creative ideas, or learning about a domain. They are also connected by design features or attributes of a generative AI application, such as the support for versioning, curation, or sandbox environments.

Our aim for these principles is threefold: (1) to provide the designers of generative AI applications with the language to discuss issues unique to \emph{generative} AI; (2) to provide strategies and guidance to help designers make important design decisions around how end users will interact with a generative AI application; and (3) to sensitize designers to the idea that generative AI applications may cause a variety of harms (likely inadvertently, but possibly intentionally). We hope these principles provide the human-AI co-creation community with a reasoned way to think through the design of novel generative AI applications.

\section{Design Principles for Generative AI Applications}
\label{sec:design-principles}

We developed seven design principles for generative AI applications based on recent research in the HCI and AI communities, specifically around human-AI co-creative processes. We conducted a literature review of research studies, guidelines, and analytic frameworks from these communities~\cite{acm2023words, amershi2019guidelines, apple, deterding2017mixed, grabe2022towards, ibm2023reid, liao2020questioning, maher2012computational, maher2022research, muller2020iccc, muller2022extending, lubart2005can, seeber2020machines, spoto2017mici}, which included experiments in human-AI co-creation~\cite{agarwal2020project, agarwal2020quality, kreminski2022evaluating, louie2020novice, sun2022investigating, weisz2021perfection, weisz2022better}, examinations of representative generative applications~\cite{Brown:GPT3, johnson2007measuring, kaiser2018generative, louie2020novice, metz:GPT-3, ramesh2022hierarchical, rombach2022high, ross2023programmers}, and a review of publications in recent workshops~\cite{geyer2021hai, muller2022genaichi, muller2022hcai, weisz2022hai}.

\subsection{The Environment: Generative Variability}
\label{sec:generative-variability}

Generative AI technologies present unique challenges for designers of AI systems compared to discriminative AI systems. First, generative AI is \emph{generative} in nature, which means their purpose is to produce artifacts as output, rather than decisions, labels, classifications, and/or decision boundaries. These artifacts may be comprised of different types of media, such as text, images, audio, animations or videos. Second, the outputs of a generative AI model are \emph{variable} in nature. Whereas discriminitive AI aims for deterministic outcomes, generative AI systems may not produce the same output for a given input each time. In fact, \emph{by design}, they can produce multiple and divergent outputs for a given input, some or all of which may be satisfactory to the user. Thus, it may be difficult for users to achieve replicable results when working with a generative AI application.

Although the very nature of generative applications violates the common HCI principle that a system should respond consistently to a user's input (for critiques of this position, see \cite{aragon2022human, boyd2012critical, costanza2020design, d2020data, gitelman2013raw, muller2022drinking}), we take the position that this environment in which generative applications operate -- \emph{generative variability} -- is a core strength. Generative applications enable users to explore or populate a ``space'' of possible outcomes to their query. Sometimes, this exploration is explicit, as in the case of systems that enable latent space manipulations of an artifact. Other times, exploration of a space occurs when a generative model produces multiple candidate outputs for a given input, such as multiple distinct images for a given prompt \cite{ramesh2022hierarchical, rombach2022high} or multiple implementations of a source code program \cite{weisz2021perfection, weisz2022better}. Recent studies also show how users may improve their knowledge of a domain by working with a generative model and its variable outputs~\cite{weisz2021perfection, ross2023programmers}.

This concept of generative variability is crucially important for designers of generative AI applications to communicate to users. Users who approach a generative AI system without understanding its probabilistic nature and its capacity to produce varied outputs will struggle to interact with it in productive ways. The design principles we outline in the following sections -- designing for multiple outcomes \& imperfection, for exploration \& human control, and for mental models \& explanations -- are all rooted in the notion that generative AI systems are distinct and unique because they operate in an environment of generative variability.

\subsection{Design for Multiple Outputs}
\label{sec:design-multiple-outputs}

Generative AI technologies such as encoder-decoder models~\cite{sutskever2014sequence, cho2014learning}, generative adversarial networks~\cite{goodfellow2020generative}, and transformer models~\cite{vaswani2017attention} are probabilistic in nature and thus are capable of producing multiple, distinct outputs for a user's input. Designers therefore need to understand the extent to which these multiple outputs should be visible to users. Do users need the ability to annotate or curate? Do they need the ability to compare or contrast? How many outputs does a user need?

Understanding the user's task can help answer these questions. If the user's task is one of \emph{production}, in which the ultimate goal is to produce a single, satisfying artifact, then designs that help the user filter and visualize differences may be preferable. For example, a software engineer's goal is often to implement a method that performs a specific behavior. Tools such as Copilot take a user's input, such as a method signature or documentation, and provide a singular output. Contrarily, if the user's task is one of \emph{exploration}, then designs that help the user curate, annotate, and mutate may be preferable. For example, a software engineer may wish to explore a space of possible test cases for a code module. Or, an artist may wish to explore different compositions or styles to see a broad range of possibilities. Below we discuss a set of strategies for helping design for multiple outputs.

\subsubsection{Versioning}
Because of the randomness involved in the generative process, as well as other user-configurable parameters (e.g. a random seed, a temperature, or other types of user controls), it may be difficult for a user to produce exactly the same outcome twice. As a user interacts with a generative AI application and creates a set of outputs, they may find that they prefer earlier outputs to later ones. How can they recover or reset the state of the system to generate such earlier outputs? One strategy is to keep track of all of these outputs, as well as the parameters that produced them, by versioning them. Such versioning can happen manually (e.g. the user clicks a button to ``save'' their current working state) or automatically.

\subsubsection{Curation}
When a generative model is capable of producing multiple outputs, users may need tools to curate those outputs. Curation may include collecting, filtering, sorting, selecting, or organizing outputs (possibly from the versioned queue) into meaningful subsets or groups, or creating prioritized lists or hierarchies of outputs according to some subjective or objective criteria. For example, CogMol\footnote{\url{http://covid19-mol.mybluemix.net}} generates novel molecular compounds, which can be sorted by various properties, such as their molecular weight, toxicity, or water solubility~\cite{chenthamarakshan2020cogmol, chenthamarakshan2020targetspecific}. In addition, the confidence of the model in each output it produced may be a useful way to sort or rank outputs, although in some cases, model confidence scores may not be indicative of the quality of the model's output~\cite{agarwal2020quality}.
    
\subsubsection{Annotation}
When a generative model has produced a large number of outputs, users may desire to add marks, decorators, or annotations to outputs of interest. These annotations may be applied to the output itself (e.g. ``I like this'') or it may be applied to a portion or subset of the output (e.g. flagging lines of source code that look problematic and need review).
    
\subsubsection{Visualizing Differences}
In some cases, a generative model may produce a diverse set of distinct outputs, such as images of cats that look strikingly different from each other. In other cases, a generative model may produce a set of outputs for which it is difficult to discern differences, such as a source code translation from one language to another. In this case, tools that aid users in visualizing the similarities and differences between multiple outputs can be useful. Depending on the users' goals, they may seek to find the \textit{invariant} aspects across outcomes, such as identifying which parts of a source code translation were the same across multiple translations, indicating a confidence in its correctness. Or, users may prioritize the \textit{variant} aspects for greater creativity and inspiration. For example, Sentient Sketchbook~\cite{liapis2013sentient} is a video game co-creation system that displays a number of different metrics of the maps it generates, enabling users to compare newly-generated maps with their current map to understand how they differ.  


\subsection{Design for Imperfection}
\label{sec:design-imperfect}

It is highly important for users to understand that the quality of a generative model's outputs will vary. Users who expect a generative AI application to produce exactly the artifact they desire will experience frustration when they work with the system and find that it often produces imperfect artifacts. By ``imperfect,'' we mean that the artifact itself may have imperfections, such as visual misrepresentations in an image, bugs or errors in source code, missing desired elements (e.g. ``an illustration of a bunny with a carrot'' fails to include a carrot), violations of constraints specified in the input prompt (e.g. ``write a 10 word sentence'' produces a much longer or shorter sentence), or even untruthful or misleading answers (e.g. a summary of a scientific topic that includes non-existent references~\cite{rose2022facebook}). But, ``imperfect'' can also mean ``doesn't satisfy the user's desire,'' such as when the user prompts a model and doesn't get back any satisfying outputs (e.g. the user didn't like any of the illustrations of a bunny with a carrot). Below we discuss a set of strategies for helping design for imperfection.

\subsubsection{Multiple Outputs}
Our previous design principle is also a strategy for handling imperfect outputs. If a generative model is allowed to produce multiple outputs, the likelihood that one of those outputs is satisfying to the user is increased. One example of this effect is in how code translation models are evaluated, via a metric called $pass@k$~\cite{roziere2020unsupervised, kulal2019spoc}. The idea is that the model is allowed to produce $k$ code translations for a given input, and if any of them pass a set of unit tests, then the model is said to have produced a correct translation. In this way, generating multiple outputs serves to mitigate the fact that the model's most-likely output may be imperfect. However, it is left up to the user to review the set of outputs and identify the one that is satisfactory; with multiple outputs that are very similar to each other, this task may be difficult~\cite{weisz2022better}, implying the need for a way to easily visualize differences.

\subsubsection{Evaluation \& Identification}
Given that generative models may not produce perfect (or perfectly satisfying) outputs, they may still be able to provide users with a signal about the quality of its output, or indicate parts that require human review. As previously discussed, a model's per-output confidence scores may be used (with care) to indicate the quality of a model's output. Or, domain-specific metrics (e.g. molecular toxicity, compiler errors) may be useful indicators to evaluate whether an artifact achieved a desirable level of quality. Thus, evaluating the quality of generated artifacts and identifying which portions of those artifacts may contain imperfections (and thus require human review, discussed further in \citet{weisz2021perfection}) can be an effective way for handling imperfection.
    
\subsubsection{Co-Creation}
User experiences that allow for co-creation, in which both the user and the AI can edit a candidate artifact, will be more effective than user experiences that assume or aim for the generative model to produce a perfect output. Allowing users to edit a model's outputs provides them with the opportunity to find and fix imperfections, and ultimately achieve a satisfactory artifact. One example of this idea is Github Copilot~\cite{web:copilot}, which is embedded in the VSCode IDE. In the case when Copilot produces an imperfect block of source code, developers are able to edit it \emph{right in context} without any friction. By contrast, tools like Midjourney or Stable Diffusion only produce a gallery of images to chose from; editing those images requires the user to shift to a different environment (e.g. Photoshop).

\subsubsection{Sandbox / Playground Environment}
A sandbox or playground environment ensures that when a user interacts with a generated artifact, their interactions (such as edits, manipulations, or annotations) do not impact the larger context or environment in which they are working. Returning to the example of Github Copilot, since it is situated inside a developer's IDE, code it produces is directly inserted into the working code file. Although this design choice enables co-creation, it also poses a risk that imperfect code is injected into a production code base. A sandbox environment that requires users to explicitly copy and paste code in order to commit it to the current working file may guard against the accidental inclusion of imperfect outputs in a larger environment or product.

\subsection{Design for Human Control}
\label{sec:design-control}

Keeping humans in control of AI systems is a core tenet of human-centered AI~\cite{shneiderman2020human, shneiderman2021human, shneiderman2022human}. Providing users with controls in generative applications can improve their experience by increasing their efficiency, comprehension, and ownership of generated outcomes \cite{louie2020novice}. But, in co-creative contexts, there are multiple ways to interpret what kinds of ``control'' people need. We identify three kinds of controls applicable to generative AI applications.

\subsubsection{Generic Controls}
One aspect of control relates to the exploration of a design space or range of possible outcomes (as discussed in Section~\ref{sec:design-explore}). Users need appropriate controls in order to drive their explorations, such as control over the number of outputs produced from an input or the amount of variability present in the outputs. We refer to these kinds of controls as \emph{generic controls}, as they are applicable to any particular generative technology or domain. As an example, some generative projects may involve a ``lifecycle'' pattern in which users benefit from seeing a great diversity of outputs in early stages of the process in order to search for ideas, inspirations, or directions. Later stages of the project may focus on a smaller number (or singular) output, requiring controls that specifically operate on that output. Many generative algorithms include a user-controllable parameter called \emph{temperature}. A low temperature setting produces outcomes that are very similar to each other; conversely, a high temperature setting produces outcomes that are very dissimilar to each other. In the ``lifecycle'' model, users may first set a high temperature for increased diversity, and then reduce it when they wish to focus on a particular area of interest in the output space. This effect was observed in a study of a music co-creation tool, in which novice users dragged temperature control sliders to the extreme ends to explore the limits of what the AI could generate~\cite{louie2020novice}.
    
\subsubsection{Technology-specific Controls}
Other types of controls will depend on the particular generative technology being employed. Encoder-decoder models, for example, often allow users to perform latent space manipulations of an artifact in order to control semantically-meaningful attributes. For example, \citet{liu2021neurosymbolic} demonstrate how semantic sliders can be used to control attributes of 3D models of animals, such as the animal's torso length, neck length, and neck rotation. Transformer models use a temperature parameter to control the amount of randomness in the generation process~\cite{vonplaten2020how}. Natural language prompting, and the emerging discipline of prompt engineering~\cite{liu2022design}, provide additional ways to tune or tweak the outputs of large language models. We refer to these kinds of controls as \emph{technology-specific controls}, as the controls exposed to a user in a user interface will depend upon the particular generative AI technology used in the application.

\subsubsection{Domain-specific Controls}
Some types of user controls will be domain-specific, dependent on the type of artifact being produced. For example, generative models that produce molecules as output might be controlled by having the user specify desired properties such as molecular weight or water solubility; these types of constraints might be propagated to the model itself (e.g. expressed as a constraint in the encoder phase), or they may simply act as a filter on the model's output (e.g. hide anything from the user that doesn't satisfy the constraints). In either case, the control itself is dependent on the fact that the model is producing a specific kind of artifact, such as a molecule, and would not logically make sense for other kinds of artifacts in other domains (e.g. how would you control the water solubility for a text-to-image model?). Thus, we refer to these types of controls, independent of how they are implemented, as \emph{domain specific}. Other examples of domain-specific controls include the reading level of a text, the color palette or artistic style of an image, or the run time or memory efficiency of source code.

\subsection{Design for Exploration}
\label{sec:design-explore}

Because users are working in an environment of generative variability, they will need some way to ``explore'' or ``navigate'' the space of potential outputs in order to identify one (or more) that satisfies their needs. Below we discuss a set of strategies for helping design for exploration.

\subsubsection{Multiple Outputs}
The ability for a generative model to produce multiple outputs (Section~\ref{sec:design-multiple-outputs}) is an enabler of exploration. Returning to the bunny and carrot example, an artist may wish to explore different illustrative styles and prompt (and re-prompt) the model for additional candidates of ``a bunny with a carrot'' in various kinds of styles or configurations. Or, a developer can explore different ways to implement an algorithm by prompting (and re-prompting) a model to produce implementations that possess different attributes (e.g. ``implement this using recursion,'' ``implement this using iteration,'' or ``implement this using memoization''). In this way, a user can get a sense of the different possibilities the model is capable of producing.

\subsubsection{Control}
Depending on the specific technical architecture used by the generative application, there may be different ways for users to control it (Section~\ref{sec:design-control}). No matter the specific mechanisms of control, \emph{providing controls} to a user provides them with the ability to interactively work with the model to explore the space of possible outputs for their given input.

\subsubsection{Sandbox / Playground Environment}
A sandbox or playground environment can enable exploration by providing a separate place in which new candidates can be explored, without interfering with a user's main working environment. For example, in a project using Copilot, \citet{cheng2022would} suggest providing, ``a sandbox mechanism to allow users to play with the prompt in the context of their own project.''

\subsubsection{Visualization}
One way to help users understand the space in which they are exploring is to explicitly visualize it for them. \citet{kreminski2022evaluating} introduce the idea of expressive range coverage analysis (ERCA) in which a user is shown a visualization of the ``range'' of possible generated artifacts across a variety of metrics. Then, as users interact with the system and produce specific artifact instances, those instances are included in the visualization to show how much of the ``range'' or ``space'' was explored by the user.

\subsection{Design for Mental Models}
\label{sec:design-mental-models}

Users form mental models when they work with technological systems \cite{scheutz2017framework, fiore2001group, mathieu2000influence}. These models represent the user's understanding of how the system works and how to work with it effectively to produce the outcomes they desire. Due to the environment of generative variability, generative AI applications will pose new challenges to users because these applications may violate existing mental models of how computing systems behave (i.e. in a deterministic fashion). Therefore, we recommend designing to support users in creating accurate mental models of generative AI applications in the following ways.

\subsubsection{Orientation to Generative Variability}
Users may need a general introduction to the concept of generative AI. They should understand that the system may produce multiple outputs for their query (Section~\ref{sec:design-multiple-outputs}), that those outputs may contain flaws or imperfections (Section~\ref{sec:design-imperfect}), and that their effort may be required to collaborate \emph{with} the system in order to produce desired artifacts via various kinds of controls (Section~\ref{sec:design-control}).
    
\subsubsection{Role of the AI}
Research in human-AI interaction suggests that users may view an AI application as filling a role such as an assistant, coach, or teammate~\cite{seeber2020machines}. In a study of video game co-creation, \citet{guzdial2019friend} found participants to ascribe roles of friend, collaborator, student, and manager to the AI system. Recent work by \citet{ross2023programmers} examined software engineers' role orientations toward a programming assistant and found that people viewed the assistant with a tool orientation, but interacted with it as if it were a social agent. Clearly establishing the role of a generative AI application in a user's workflow, as well as its level of autonomy (e.g. \cite{fitts1951human, sheridan1978human, parasuraman2000model, horvitz1999principles}), will help users better understand how to interact effectively with it. Designers can reason about the role of their application by answering questions such as, is it a tool or partner? does it act proactively or does it just respond to the user? does it make changes to an artifact directly or does it simply make recommendations for the user?

\subsection{Design for Explanations}
\label{sec:design-xai}

Generative AI applications will be unfamiliar and possibly unusual to many users. They will want to know what the application can (and cannot) do, how well it works, and how to work with it effectively. Some users may even wish to understand the technical details of how the underlying generative AI algorithms work, although these details may not be necessary to work effectively with the model (as discussed in \cite{weisz2021perfection}).

In recent years, the explainable AI (XAI) community has made tremendous progress at developing techniques for explaining how AI systems work~\cite{arya2020ai, ehsan2022human, liao2020questioning, liao2021introduction, simkute2022xai}. Much of the work in XAI has focused on discriminative algorithms: how they generally make decisions (e.g. via interpretable models~\cite[Chapter 5]{molnar2020interpretable} or feature importance~\cite[Section 8.5]{molnar2020interpretable}, and why they make a decision in a specific instance (e.g. via counterfactual explanations~\cite[Section 9.3]{molnar2020interpretable}.

Recent work in human-centered XAI (HCXAI) has emphasized designing explanations that cater to human knowledge and human needs~\cite{ehsan2022human}. This work grew out of a general shift toward human-centered data science~\cite{aragon2022human}, in which the import of explanations is not for a technical user (data scientist), but for an end user who might be impacted by a machine learning model.

In the case of generative AI, recent work has begun to explore the needs for explainability. \citet{sun2022investigating} explored explainability needs of software engineers working with a generative AI model for various types of use cases, such as code translation and autocompletion. They identified a number of types of questions that software engineers had about the generative AI, its capabilities, and its limitations, indicating that explainability is an important feature for generative AI applications. They also identified several gaps in existing explainability frameworks stemming from the \emph{generative} nature of the AI system, indicating that existing XAI techniques may not be sufficient for generative AI applications. Thus, we make the following recommendations for how to design for explanations.

\subsubsection{Calibrate Trust by Communicating Capabilities and Limitations}
Because of the inherent imperfection of generative AI outputs, users would be well-served if they understood the limitations of these systems~\cite{muller2022forgetting, pinanez2021expose}, allowing them to \emph{calibrate} their trust in terms of what the application can and cannot do~\cite{pinanez2022breakdowns}. When these kinds of imperfections (Section~\ref{sec:design-imperfect}) are not signaled, users of co-creative tools may mistakenly blame themselves for shortcomings of generated artifacts in co-creative applications ~\cite{louie2020novice}, and users in Q \& A use cases can be shown deceptive misconceptions and harmful falsehoods as objective answers \cite{lin2021truthfulqa}. One way to communicate the capabilities of a generative AI application is to show examples of what it can do. For example, Midjourney provides a public discussion space to orient new users and show them what other users have produced with the model. This space not only shows the outputs of the model (e.g. images), but the textual prompts that produced the images. In this way, users can more quickly come to understand how different prompts influence the application's output. To communicate limitations, systems like ChatGPT contain modal screens to inform users of the system's limitations.

\subsubsection{Use Explanations to Create and Reinforce Accurate Mental Models}
\citet{weisz2021perfection} explored how a generative model's confidence could be surfaced in a user interface. Working with a transformer model on a code translation task, they developed a prototype UI that highlighted tokens in the translation that the model was not confident in. In their user study, they found that those highlights also served as explanations for how the model worked: users came to understand that each source code token was chosen probabilistically, and that the model had considered other alternatives. This design transformed an algorithmic weakness (imperfect output) into a resource for users to understand how the algorithm worked, and ultimately, to control its output (by showing users where they might need to make changes).

\subsection{Design Against Harms}
\label{sec:design-against-harms}

The use of AI systems -- including generative AI applications -- may unfortunately lead to diverse forms of harms, especially for people in vulnerable situations. Much work in AI ethics communities has identified how discriminative AI systems may perpetuate harms such as the denial of personhood or identity~\cite{costanza2020design, kantayya2020coded, spiel2021they}; the deprivation of liberty or children \cite{lyn2020risky, saxena2021framework}, and the erasure of persons, cultures, or nations through data silences \cite{muller2022forgetting}. We identify four types of potential harms, some of which are unique to the generative domain, and others which represent existing risks of AI applications that may manifest in new ways.

Our aim in this section is to sensitize designers to the potential risks and harms that generative AI systems may pose. We do not prescribe solutions to address these risks, in part because it is an active area of research to understand how these kinds of risks could be mitigated. Risk identification, assessment, and mitigation is a sociotechnical problem involving computing resources, humans, and cultures. Even with our focus on the design of generative applications, an analysis of harms that is limited to design concepts may blur into technosolutionism~\cite{lindtner2016reconstituting, madaio2020co, resseguier2021ethics}.

We do posit that \emph{human-centered} approaches to generative AI design are a useful first step, but must be part of a larger strategy to understand who are the direct and indirect stakeholders of a generative application~\cite{friedman2019value, hendry2021value}, and to work directly with those stakeholders to \emph{identify} harms, \emph{understand} what are their differing priorities and value tensions~\cite{miller2007value}, and \emph{negotiate} issues of culture, policy, and (yes) technology to meet these diverse challenges (e.g., \cite{denzin2008handbook, disalvo2022design, hayes2014knowing}).

\subsubsection{Hazardous Model Outputs}

Generative AI applications may produce artifacts that cause harm. In an integrative survey paper, \citet{weidinger2021ethical} list six types of potential harms of large language models, three of which regard the harms that may be caused by the model's output: 

\begin{itemize}
    \item \textbf{Discrimination, Exclusion, and Toxicity}. Generative models may produce outputs that promote discrimination against certain groups, exclude certain groups from representation, or produce toxic content. Examples include text-to-image models that fail to produce ethnically diverse outputs for a given input (e.g. a request for images of doctors produces images of male, white doctors \cite{cho2022dall} or language models that produce inappropriate language such as swear words, hate speech, or offensive content~\cite{acm2023words, ibm2023reid}.
    
    \item \textbf{Information Hazards}. Generative models may inadvertently leak private or sensitive information from their training data. For example,  \citet{carlini2021extracting} found that strategically prompting GPT-2 revealed an individual's full name, work address, phone number, email, and fax number. Additionally, larger models may be more vulnerable to these types of attacks~\cite{carlini2021extracting, carlini2022quantifying}.
    
    \item \textbf{Misinformation Harms}. Generative models may produce inaccurate misinformation in response to a user's query. Lin et al. \cite{lin2021truthfulqa} found that GPT-3 can provide false answers that mimic human falsehoods and misconceptions, such as ``coughing can help stop a heart attack'' or ``[cold weather] tells us that global warming is a hoax''. \citet{singhal2022large} caution against the tendency of LLMs to hallucinate references, especially if consulted for medical decisions. \citet{albrecht2022despite} claim that LLMs have few defenses against adversarial attacks while advising about ethical questions. The Galactica model was found to hallucinate non-existent scientific references~\cite{heaven_2022}, and Stack Overflow has banned responses sourced from ChatGPT due to their high rate of incorrect, yet plausible, responses~\cite{vincent_2022}.
\end{itemize}

In addition to those harms, a generative model's outputs may be hazardous in other ways as well.

\begin{itemize}
    \item \textbf{Deceit, Impersonation, and Manipulation}. Generative algorithms can be used to create false records or ``deep fakes'' (e.g., \cite{houde2020business, meskys2020regulating}), to impersonate others (e.g. \cite{stupp2019fraudsters}), or to distort information into politically-altered content~\cite{yang2019unsupervised}. In addition, they may manipulate users who believe that they are chatting with another human rather than with an algorithm, as in the case of an unreviewed ChatGPT ``experiment'' in which at least 4,000 people seeking mental health support were connected to a chatbot rather than a human counselor~\cite{morris2023we}.

    \item \textbf{Copyright, Licenses, and Intellectual Property}. Generative models may have been trained on data protected by regulations such as the GDPR, which prohibits the re-use of data beyond the purposes for which it was collected. In addition, large language models have been referred to as ``stochastic parrots'' due to their ability to reproduce data that was used during their training~\cite{bender2021dangers}. One consequence of this effect is that the model may produce outputs that incorporate or remix materials that are subject to copyright or intellectual property protections~\cite{franceschelli2022copyright, hristov2016artificial, murray2022generative}. For example, the Codex model, which produces source code as output, may (re-)produce source code that is copyrighted or subject to a software license, or that was openly shared under a creative commons license that prohibits commercial re-use (e.g., in a pay-to-access LLM). Thus, the use of a model's outputs in a project may cause that project to violate copyright protections, or subject that project to a restrictive license (e.g. GPL). As of this writing, there is a lawsuit against GitHub, Microsoft, and OpenAI on alleged copyright violations in the training of Codex~\cite{web:githubcopilotlitigation}.
\end{itemize}

\subsubsection{Misuse}

\citet{weidinger2021ethical} describe how generative AI applications may be misused in ways unanticipated by the creators of those systems. Examples include making disinformation cheaper and more effective, facilitating fraud and scams, assisting code generation for cyberattacks, or conducting illegitimate surveillance and censorship. In addition to these misuses, \citet{houde2020business} also identify business misuses of generative AI applications such as facilitating insurance fraud and fabricating evidence of a crime. Although designers may not be able to prevent users from intentionally misusing their generative AI applications, there may be preventative measures that make sense for a given application domain. For example, output images may be watermarked to indicate they were generated by a particular model, blocklists may be used to disallow undesirable words in a textual prompt, or multiple people may be required to review or approve a model's outputs before they can be used.

\subsubsection{Human Displacement}

One consequence of the large-scale deployment of generative AI technologies is that they may come to \emph{replace}, rather than \emph{augment} human workers. Such concerns have been raised in related areas, such as the use of automated AI technologies in data science~\citet{wang2019human, wang2021towards}. \citet{weidinger2021ethical} specifically discuss the potential economic harms and inequalities that may arise as a consequence of widespread adoption of generative AI. If a generative model is capable of producing high-fidelity outputs that rival (or even surpass) what can be created by human effort, are the humans necessary anymore? Contemporary fears of human displacement by generative technologies are beginning to manifest in mainstream media, such as in the case of illustrators' concerns that text-to-image models such as Stable Diffusion and Midjourney will put them out of a job~\cite{wilkins2022will}. We urge designers to find ways to design generative AI applications that \emph{enhance} or \emph{augment} human abilities, rather than applications that aim to \emph{replace} human workers. Copilot serves as one example of a tool that clearly enhances the abilities of a software engineer: it operates on the low-level details of a source code implementation, freeing up software engineers to focus more of their attention on higher-level architectural and system design issues.

\section{Discussion}
\label{sec:discussion}

\subsection{Designing for User Aims}

Users of generative AI applications may have varied aims or goals in using those systems. Some users may be in pursuit of \emph{perfecting a singular artifact}, such as a method implementation in a software program. Other users may be in \emph{pursuit of inspiration or creative ideas}, such as when exploring a visual design space. As a consequence of working with a generative AI application, users may also \emph{enhance their own learning or understanding} of the domain in which they are operating, such as when a software engineer learns something new about a programming language from the model's output. Each of these aims can be supported by our design principles, as well as help designers determine the appropriate strategy for addressing the challenges posed by each principle.

To support artifact production, designers ought to carefully consider how to manage a model's multiple, imperfect outputs. Interfaces ought to support users in curating, annotating, and mutating artifacts to help users refine a singular artifact. The ability to version artifacts, or show a history of artifact edits, may also be useful to enable users to revisit discarded options or undo undesirable modifications. For cases in which users seek to produce one ``ideal'' artifact that satisfies some criteria, controls that enable them to co-create with the generative tool can help them achieve their goal more efficiently, and explanations that signal or identify imperfections can tell them how close or far they are from the mark.

To support inspiration and creativity, designers also ought to provide adequate controls that enable users to explore a design space of possibilities~\cite{kreminski2022evaluating, morris2022design}. Visualizations that represent the design space can also be helpful as they can show which parts the user has vs. has not explored, enabling them to explore the novel parts of that space. Tools that help users manage, curate, and filter the different outputs created during their explorations can be extremely helpful, such as a digital mood board for capturing inspiring model outputs.

Finally, to support learning how to effectively interact with a generative AI application, designers ought to help users create accurate mental models~\cite{kollmansberger2010helping} through explanations~\cite{arya2020ai, ehsan2022human, liao2020questioning, liao2021introduction, simkute2022xai}. Explanations can help answer general questions such as what a generative AI application is capable or not capable of generating, how the model's controls impact its output, and how the model was trained and the provenance of its training data. They can also answer questions about a specific model output, such as how confident the model was in that output, which portions of that output might need human review or revision, how to adjust or modify the input or prompt to adjust properties of the output, or what other options or alternatives exist for that output.

\subsection{The Importance of Value-Sensitive Design in Mitigating Potential Harms}

Designers need to be sensitive to the potential harms that may be caused by the rapid maturation and widespread adoption of generative AI technologies. Although sociotechnical means for mitigating these harms have yet to be developed, we recommend that designers use a Value Sensitive Design approach~\cite{friedman2019value, hendry2021value} when reasoning about how to design generative AI applications. By clearly identifying the different stakeholders and impacted parties of a generative AI application, and explicitly enumerating their values, designers can make more reasoned judgments about how those stakeholders might be impacted by hazardous model outputs, model misuse, and issues of human displacement.

\section{Limitations and Future Work}

Generative AI applications are still in their infancy, and new kinds of co-creative user experiences are emerging at a rapid pace. Thus, we consider these principles to be in their infancy as well, and it is possible that other important design principles, strategies, and/or user aims have been overlooked. In addition, although these principles can provide helpful guidance to designers in making specific design decisions, they need to be validated in real-world settings to ensure their clarity and utility.

\section{Conclusion}

We present a set of seven design principles for generative AI applications. These principles are grounded in an environment of generative variability, the key characteristics of which are that a generative AI application will generate artifacts as outputs, and those outputs may be varied in nature (e.g. of varied quality or character). The principles focus on designing for multiple outputs and the imperfection of those outputs, designing for exploration of a space or range of possible outputs and maintaining human control over that exploration, and designing to establish accurate mental models of the generative AI application via explanations. We also urge designers to design \emph{against} the potential harms that may be caused by hazardous model output (e.g. the production of inappropriate language or imagery, the reinforcement of existing stereotypes, or a failure to inclusively represent different groups), by misuse of the model (e.g. by creating disinformation or fabricating evidence), or by displacing human workers (e.g. by designing for the \emph{replacement} rather than the \emph{augmentation} of human workers). We envision these principles to help designers make reasoned choices as they create novel generative AI applications.

\bibliographystyle{ACM-Reference-Format}
\bibliography{references}


\end{document}